\documentclass[prd,amsmath,amssymb,floatfix,superscriptaddress,notitlepage,nofootinbib,preprintnumbers,twocolumn]{revtex4-1}
\usepackage{amsfonts,amssymb,amsmath,graphicx,color,bm,enumitem}
\usepackage{hyperref}
\usepackage[utf8]{inputenc}
\usepackage{amsthm}
\usepackage[english]{babel}
\definecolor{ultramarine}{rgb}{0.07, 0.04, 0.56}
\definecolor{cadmiumgreen}{rgb}{0.0, 0.42, 0.24}
\definecolor{indigo(dye)}{rgb}{0.0, 0.25, 0.42}
\hypersetup{
colorlinks=true,
citecolor=blue,
linkcolor=green,
urlcolor=indigo(dye),
}

\newcommand{\be}{\begin{eqnarray}}  
\newcommand{\ee}{\end{eqnarray}}
\newcommand{\bem}{\begin{pmatrix}}
\newcommand{\eem}{\end{pmatrix}}

\newcommand{\Q}{\mathcal{Q}}

\newcommand{\X}{\mathcal{X}}
\newcommand{\Y}{\mathcal{Y}}

\begin{document}

\title{
Black holes in the quadratic-order extended vector-tensor theories
}

\author{Masato Minamitsuji}
\affiliation{Centro de Astrof\'{\i}sica e Gravita\c c\~ao  - CENTRA, Departamento de F\'{\i}sica, Instituto Superior T\'ecnico - IST, Universidade de Lisboa - UL, Av. Rovisco Pais 1, 1049-001 Lisboa, Portugal}

\begin{abstract}
We investigate the static and spherically black hole solutions in the quadratic-order extended vector-tensor theories without suffering from the Ostrogradsky instabilities, which include the quartic-order (beyond-)generalized Proca theories as the subclass. We start from the most general action of the vector-tensor theories constructed with up to the quadratic-order terms of the first-order covariant derivatives of the vector field, and derive the Euler-Lagrange equations for the metric and vector field variables in the static and spherically symmetric backgrounds. We then substitute the spacetime metric functions of the Schwarzschild, Schwarzschild-de Sitter/ anti-de Sitter,   Reissner-Nordstr\"{o}m-type, and Reissner-Nordstr\"{o}m-de Sitter/ anti-de Sitter-type solutions and the vector field with the constant spacetime norm into the Euler-Lagrange equations, and obtain the conditions for the existence of these black hole solutions. These solutions are classified into the two cases 1) the solutions with the vanishing vector field strength; the stealth Schwarzschild and the Schwarzschild de Sitter/ anti- de Sitter solutions, and 2) those with the nonvanishing vector field strength; the charged stealth Schwarzschild and the charged Schwarzschild de Sitter/ anti- de Sitter solutions, in the case that the tuning relation among the coupling functions is satisfied. In the latter case, if this tuning relation is violated, the solution becomes the Reissner-Nordstr\"{o}m-type solution. We show that the conditions for the existence of these solutions are compatible with the degeneracy conditions for the Class-A theories, and recover the black hole solutions in the generalized Proca theories as the particular cases.
\end{abstract}
\maketitle

\section{Introduction}
\label{sec1}

\subsection{Black holes in modified theories of gravitation}
\label{sec1a}

Black holes are the most fundamental objects  
not only in general relativity 
but also in many theories of gravitation,
and hence 
the properties of black hole solutions
will provide us opportunities
to test modified theories of gravitation
from the theoretical and observational viewpoints.
It is well known that 
many theories of gravitation 
share the same black hole solutions 
with general relativity (GR)
\cite{Psaltis:2007cw,Motohashi:2018wdq}.
In vacuum GR, there is the uniqueness theorem stating that the asymptotically flat 
and stationary black holes solutions are described only by three parameters, 
i.e., mass, electric charge, and angular momentum \cite{Israel:1967wq,nh1,nh3,nh5}. 
The no-hair BH theorem is valid for the canonical scalar field minimally coupled to gravity
\cite{Chase:1970,nh4},
and also holds for the scalar-tensor theories 
with 
the direct coupling of the scalar field to the Ricci scalar \cite{nh2,nh6,nh7},
the noncanonical kinetic terms of the scalar field \cite{nh10,nh11}
and 
the higher-order derivative interactions of the scalar field~\cite{nh8,nh9}.
These no-hair theorems state 
that the Schwarzschild or Kerr solution
with the vanishing scalar or vector field
is the unique vacuum black hole solution
under the given symmetries of the spacetime.
On the other hand,
the black hole solutions
with the nontrivial profiles of the scalar and vector fields
have also been found 
in the scalar-tensor and vector-tensor theories
which are free from some of the assumptions for the no-hair theorems,
for example,
via the conformal coupling to the Ricici scalar with the singular behavior of 
the scalar field at the event horizon \cite{bbmb1,bbmb2},
the nonminimal coupling to the Gauss-Bonnet term
\cite{edgb1,edgb2,edgb3,edgb4,sz1,sz2},
the linear time dependence of the scalar field
in the shift-symmetric scalar-tensor theories
\cite{Mukohyama:2005rw,Babichev:2013cya,Appleby:2015ysa,Babichev:2016rlq,bhbh1,bhbh2,Babichev:2016kdt,dhostbh1,dhostbh2,dhostbh3,dhostbh4,dhostbh5},
the asymptotically anti-de Sitter spacetimes \cite{AyonBeato:2004ig,Rinaldi:2012vy,Anabalon:2013oea,Minamitsuji:2013ura},
and 
the self-derivative couplings
and 
the nonminimal couplings of the vector field to the spacetime curvature
\cite{Chagoya:2016aar,Minamitsuji:2016ydr,Babichev:2017rti,Heisenberg:2017xda,Heisenberg:2017hwb,Minamitsuji:2017aan,Kase:2018owh}
(see also Ref. \cite{Herdeiro:2015waa} for a review).

\subsection{The quadratic-order extended vector-tensor theories}
\label{sec1b}

In this paper, 
we will investigate the static and spherically 
black hole solutions
with the constant norm of the vector field
in the quadratic-order extended vector-tensor theories \cite{Kimura:2016rzw},
which are currently recognized as the 
most general single-field vector-tensor theories
without the Ostrogradsky instabilities \cite{Woodard:2015zca}.
The construction of the theories
first considers the most general vector-tensor theories,
which are constructed with up 
to the quadratic-order terms of 
the first-order covariant derivatives of the vector field,
and imposes the degeneracy conditions
to avoid the appearance of the Ostrogradsky instabilities (see below).
The quadratic-order extended vector-theories also
correspond to the extension of the quadratic-order degenerate 
higher-order scalar-tensor (DHOST) theories 
\cite{Langlois:2015cwa,Achour:2016rkg,BenAchour:2016fzp,Langlois:2018dxi},
which are known as the most general single-field scalar-tensor theories
without suffering from the Ostrogradsky instabilities \cite{Woodard:2015zca},
to the vector-tensor theories.
The extension is similar to that  
from the Horndesky theories \cite{Horndeski:1974wa,Deffayet:2009mn,Kobayashi:2011nu}
to the generalized Proca theories
\cite{Tasinato:2014eka,Heisenberg:2014rta,DeFelice:2016cri,Heisenberg:2018vsk}.

The most general vector-tensor theories
which are constructed with up to the quadratic-order terms 
of the first-order covariant derivatives
of the vector field $\nabla_\mu A_\nu$
are given by 
\begin{eqnarray}
\label{action}
S
=
\int  d^4x\sqrt{-g}
\left[
f_0 (\Y)
+
f_2\left({\Y} \right)
R
+{C}^{\mu\nu\rho\sigma}
{\nabla}_\mu A_\nu 
{\nabla}_\rho A_\sigma
\right],
\nonumber\\
&&
\end{eqnarray}
\cite{Kimura:2016rzw},
where
the Greek indices $(\mu, \nu,\cdots)$
run the four-dimensional spacetime,
$g_{\mu\nu}$ represents the metric tensor,
$\nabla_\mu$ and $R$
are the covariant derivative and Ricci scalar
associated with $g_{\mu\nu}$,
respectively, 
$A_\mu$ is the vector field,
\be
\label{norm}
\Y:=g^{\mu\nu}A_\mu A_\nu,
\ee
is the spacetime norm of the vector field $A_\mu$,
and 
$f_0(\Y)$ and $f_2(\Y)$ are the free functions of $\Y$,
respectively.
We also define the symmetric rank-4 tensor
constructed with the inverse metric and the vector field
\cite{Kimura:2016rzw};
\begin{eqnarray}
\label{defc4}
C^{\mu\nu\rho\sigma}
&:=&
\alpha_1 ({\cal Y})
g^{\mu (\rho} g^{\sigma)\nu}
+
\alpha_2 ({\cal Y})
g^{\mu\nu}g^{\rho\sigma}
\nonumber\\
&+&
\frac{\alpha_3 ({\cal Y})}{2}
\left(
   A^\mu A^\nu g^{\rho\sigma}
+ A^\rho A^\sigma g^{\mu\nu} 
\right)
\nonumber\\
&+&
\frac{\alpha_4 ({\cal Y})}{2}
\left(
   A^\mu A^{(\rho} g^{\sigma)\nu}
+
   A^\nu A^{(\rho} g^{\sigma)\mu}
\right)
\nonumber\\
&+&
\alpha_5({\cal Y})
A^\mu A^\nu A^\rho A^\sigma
+
\alpha_6 ({\cal Y})
g^{\mu[\rho} g^{\sigma]\nu}
\nonumber\\
&+&
\frac{\alpha_7 ({\cal Y})}{2}
\left(
     A^\mu A^{[\rho} g^{\sigma]\nu}
  - A^\nu A^{[\rho} g^{\sigma]\mu}
\right)
\nonumber\\
&+&
\frac{\alpha_8({\cal Y})}{4}
\left(
  A^\mu A^{\rho} g^{\sigma\nu}
-A^\nu A^{\sigma} g^{\mu\rho}
\right),
\end{eqnarray}
where 
$\alpha_i(\Y)$ (the Latin induces run $i=1,2,\cdots, 7, 8$)
are also the free functions of $\Y$.
By introducing the symmetric and antisymmetric parts
of the first-order covariant derivative of the vector field $\nabla_\mu A_\nu$,
respectively,
\begin{subequations}
\be
&&
S_{\mu\nu}:=\nabla_\mu A_\nu+\nabla_\nu A_\mu,
\\
&&
F_{\mu\nu}:=
\nabla_\mu A_\nu- \nabla_\nu A_\mu
\left(
=
\partial_\mu A_\nu- \partial_\nu A_\mu
\right),
\ee
\end{subequations}
the $C^{\mu\nu\rho\sigma}$ term in the action \eqref{action}
with Eq. \eqref{defc4}
can be rewritten as 
\begin{eqnarray}
&&
4C^{\mu\nu\rho\sigma}
\nabla_\mu A_\nu 
\nabla_\rho A_\sigma
\nonumber\\
&=&
 \alpha_1(\Y)
   S_{\mu\nu}S^{\mu\nu}
+\alpha_2(\Y)
   \left(S^\mu{}_\mu\right)^2
\nonumber\\
&+&\alpha_3(\Y)
  A^\mu A^\nu S_{\mu\nu} S_\rho{}^\rho
+
\alpha_4(\Y)
 A^\mu A^\nu S_{\mu\rho}S_\nu{}^\rho
\nonumber\\
&+&
\alpha_5(\Y)
 \left(A^\mu A^\nu S_{\mu\nu}\right)^2
+
\alpha_6(\Y)
  F_{\mu\nu}F^{\mu\nu}
\nonumber\\
&+&
\alpha_7(\Y)
 A^\mu A^\nu F_{\mu\rho} F_{\nu} {}^\rho
+\alpha_8 (\Y)
A^\mu A^\nu F_{\mu}{}_\rho S_{\nu\rho}.
\end{eqnarray}
In general,
the vector-tensor theories \eqref{action} with Eq. \eqref{defc4}
are not free from the Ostrogradsky instabilities \cite{Woodard:2015zca}.
The key idea to avoid the appearance of the Ostrogradsky instabilities
is to impose the degeneracy conditions 
among the highest-derivative equations of motion.

Here, we demonstrate
how the degeneracy condition can reduce the higher-derivative system 
to the second-order system 
with a simple example in the context of analytical mechanics
\cite{Langlois:2015cwa,Langlois:2018dxi,Kobayashi:2019hrl,Motohashi:2016ftl}.
We consider a particle 
following the trajectory $\left(x(T), y(T)\right)$
that minimizes the action $S_p=\int  dT L_p$;
\be
\label{toy}
L_p=\frac{a_1}{2}\ddot{x}^2
   + a_2 \ddot{x} \dot{y}
    +\frac{a_3}{2} \dot{y}^2
   + \frac{1}{2}\dot{x}^2
-V(x,y),
\ee
where
$T$ repesents the time coordinate
\footnote{
In order to avoid any confusion, we 
would like to distuinguish the time coordinate $T$
from that in the static and spherically symmetric spacetime $t$.
(see Eq. \eqref{ansatza}). 
$T$ here represents the time coordinate in the mechanical system
for a point particle following the trajectory $\left(x(T), y(T)\right)$.},
`dot' means the derivative with respect to $T$,
$a_j$ ($j=1,2,3$) are constants,
and 
$V(x,y)$ represents the potential term.
The Euler-Lagrange equations for $x$ and $y$ 
are given by the fourth-order and second-order 
differential equations,
and hence the theory generically contains
the three degrees of freedom,
one of which corresponds to the Ostrogradsky ghost.  
By eliminating $y$ with $y=z-(a_2/a_3)\dot{x}$,
the above Lagrangian reduces to 
\be
L_p=\frac{1}{2}\left(a_1-\frac{a_2^2}{a_3}\right)\ddot{x}^2
+ \frac{a_3}{2} \dot{z}^2
+\frac{1}{2} \dot{x}^2
-V(x,y).
\ee
Hence, 
by imposing $a_2^2-a_1a_3=0$,
which is called the {\it degeneracy condition},
the highest derivative term for $x$ can be eliminated,
and 
the system reduces to the second-order system of $(x,z)$
and contains only two physical degrees of freedom,
namely, the Ostrogradsky ghost is eliminated.
In Appendix \ref{app_a},
we will review
the theory \eqref{toy} in terms of the Hamiltonian analysis,
and
that the degeneracy condition makes 
the Hamiltonian bounded from below.
We note that more general class of the degenerate higher-derivative theories
in analytical mechanics,
has been investigated
in Refs. \cite{Motohashi:2014opa,Motohashi:2016ftl,Klein:2016aiq,Motohashi:2017eya,Motohashi:2018pxg}.

The {\it degenerate} theories
have been extended
from analytical mechanics to scalar-tensor theories
in Refs.~\cite{Langlois:2015cwa,Achour:2016rkg,BenAchour:2016fzp,Langlois:2018dxi}
and to vector-tensor theories in Ref. \cite{Kimura:2016rzw}.
In the case of the vector-tensor theories,
after the Arnowitt-Deser-Misner  (ADM) decomposition \cite{Arnowitt:1959ah}
of  the theory \eqref{action} with Eq. \eqref{defc4},
the degeneracy conditions 
yield the three solutions,
namely,
the three different classes of the degenerate theories \cite{Kimura:2016rzw},
which are briefly summarized as follows
(see Sec. \ref{sec4} for details)
\footnote{In this paper, we call
`Case A-C' in Ref. \cite{Kimura:2016rzw} `Class A-C', respectively.}
;

\begin{itemize}

\item
Class A; $\alpha_1+\alpha_2=0$ and $f_2\neq 0$,
\label{ca}

\item
Class B; $\alpha_1+\alpha_2\neq 0$ and $f_2\neq 0$ ,
\label{cb}

\item 
Class C;  $f_2=0$.
\label{cc}

\end{itemize}

Class A
includes the quadratic- and quartic-order (beyond-)generalized Proca theories
as the particular subclasses
and can be mapped from them via the vector disformal transformation,
while 
Classes B and C
cannot be related to the (beyond-)generalized Proca theories via the same transformation.
Within Class A,
there are the subclasses from Class A1 to Class A4 \cite{Kimura:2016rzw},
which will be explained in subsection \ref{sec4a}.
In  Class B,
there are the subclasses from Class B1 to Class B6 \cite{Kimura:2016rzw},
which we omit to show explicitly.
We exclude Class C from our analysis,
since there is no Einstein-Hilbert term 
in the gravitational action
and hence no way to compare with the results in general relativity.
Thus, 
in the rest of this paper, 
we focus on Class A and Class B.

As in the case of the quadratic-order DHOST theories  
\cite{dhostbh1,dhostbh2,dhostbh3,dhostbh4,dhostbh5},
while the degeneracy conditions themselves 
are not relevant for the derivation of the static black hole solutions
in the vector-tensor theories (see Secs. \ref{sec2} and \ref{sec3}),
we will check the compatibility of our conditions 
with the above three classes
in the quadratic-order extended vector-tensor theories
(see Sec. \ref{sec4}).
It should be noted that  
the quadratic-order extended vector-tensor theories
reduce to the quadratic-order DHOST theories with the shift symmetry
in the limit of $A_\mu\to \partial_\mu \phi$,
i.e., the field strength of the vector field vanishes,
where $\phi$ corresponds to the scalar field \cite{Kimura:2016rzw}
(see Appendix \ref{app_b}).

\subsection{The static and spherically symmetric black hole solutions}
\label{sec1c}

In the vector-tensor theories without the $U(1)$ gauge symmetry,
the hairy black hole solutions 
have been investigated 
in Refs. 
\cite{Chagoya:2016aar,Minamitsuji:2016ydr,Babichev:2017rti,Heisenberg:2017xda,Heisenberg:2017hwb,Minamitsuji:2017aan}.
One of the interesting solutions among them is
of  {\it stealth} type,
which describes the Schwarzschild or Kerr black hole solutions
in the scalar-tensor or vector-tensor theories
where the scalar or vector field with the nontrivial profile
does not backreact on the spacetime geometry.
The steath black hole solutions
have been explored in Refs.~\cite{AyonBeato:2004ig,Mukohyama:2005rw,Babichev:2013cya,Chagoya:2016aar,Takahashi:2020hso}.
Since the metric functions do not explicitly depend on the model parameters,
these black hole solutions cannot be distinguished from those in GR,
unless the scalar or vector field has the direct coupling to the matter sector.
The stealth Schwarzschild solution has been obtained 
at the first time
in the class of 
the generalized Proca theories with the nonminimal coupling 
to the Einstein tensor $G^{\mu\nu}A_\mu A_\nu$,
where $G_{\mu\nu}$ represents the Einstein tensor \cite{Chagoya:2016aar}.
The Schwarzschild de Sitter/ andi- de Sitter
and Reissner-Nordstr\"{o}m de Sitter/ anti-de Sitter solutions
have been obtained
in the class of the generalized Proca theories with the nonminimal coupling $G^{\mu\nu}A_\mu A_\nu$
and the mass term $m^2 g^{\mu\nu}A_\mu A_\nu$
in Ref. \cite{Minamitsuji:2016ydr},
where the mass parameter $m$ plays the role of the effective cosmological constant
which is different from the bare cosmological constant $\Lambda_b$.
These analytic solutions have been found
in the other classes of the generalized Proca theories 
\cite{Babichev:2017rti,Heisenberg:2017xda,Heisenberg:2017hwb,Minamitsuji:2017aan}.
While these analytic black hole solutions
have been obtained under the assumption
that the vector field has a constant spacetime norm $\Y={\rm const}$,
in the case that $\Y$ is not a constant 
black hole solutions with the nontrivial vector field profile
have been investigated numerically in Ref.
\cite{Babichev:2017rti,Heisenberg:2017xda,Heisenberg:2017hwb}.
These works have been extended to the case of 
the beyond-generalized Proca theories in Ref. \cite{Kase:2018owh}.
The analysis in the present paper 
will analyze the static and spherically symmetric solutions 
with the constant spacetime norm of the vector field $\Y={\rm const}$
in the quadratic-order extended vector-tensor theories \eqref{action}.
Needless to say,
our work corresponds to
the direct extension of the above previous works
\cite{Chagoya:2016aar,Minamitsuji:2016ydr,Minamitsuji:2017aan}.

Our black hole solutions
will also be the extension of the works
on hairy static and spherically symmetric 
black hole solutions in 
the shift-symmetric quadratic-order DHOST theories
\cite{dhostbh1,dhostbh2,dhostbh3,dhostbh4}
to the quadratic-order extended vector-tensor theories.
The conditions for the existence of 
the stealth Schwarzschild
and Schwarzschild de Sitter/anti- de Sitter solutions
in the quadratic-order DHOST theories
will be reviewed in Appendix \ref{app_b}.
As seen in Appendix \ref{app_b},
the several classes of the solutions 
obtained in Secs. \ref{sec2} and \ref{sec3}
with the vanishing electric field strength
will reproduce
those for the static and spherically symmetric 
hairy black hole solutions
in the shift-symmetric quadratic-order DHOST theories
in the limit $A_\mu\to \partial_\mu\phi$.
In other words,
the investigation of the black hole solutions
in the quadratic-order extended vector-tensor theories
will be along these previous studies,
and 
should enrich our knowledge on the black hole solutions
in modified theories of gravitation
and clarify the relationship between different theories
in terms of black hole physics.
This is the main motivation of our work.

Along Refs. \cite{Chagoya:2016aar,Minamitsuji:2016ydr,Heisenberg:2017xda,Heisenberg:2017hwb,Minamitsuji:2017aan},
we investigate the exact, static and spherically symmetric vacuum solutions
in the quadratic-order extended vector-tensor theories.
In order to obtain the exact black hole solutions, 
we follow the same strategy 
as that in the case of the quadratic-order DHOST theories
\cite{dhostbh1,dhostbh2,dhostbh3}.
First,
we restrict our attention 
on the static and spherically symmetric solutions
with the following ansatz;
\begin{subequations}
\label{ansatz}
\be
\label{ansatza}
ds^2
&=&g_{\mu\nu}
  dx^\mu dx^\nu
\nonumber\\
&=&
-f(r)dt^2
+\frac{dr^2}{h(r)}
+
r^2 
\left(
  d\theta^2
+\sin^2\theta 
  d\varphi^2
\right),
\\
\label{ansatzb}
A_\mu dx^\mu
&=&
 A_t(r) dt
+A_r(r) dr,
\ee
\end{subequations}
with 
\be
\Y=-\frac{A_t(r)^2}{f(r)} +A_r(r)^2 h(r),
\ee
from Eq. \eqref{norm},
where $t$, $r$, and $(\theta,\varphi)$
represent the time, radial, and angular coordinates, respectively.
$f(r)$, $h(r)$, $A_t(r)$, and $A_r(r)$
are the functions of $r$.
Substituting Eq.~\eqref{ansatz} into 
the action \eqref{action} with Eq.~\eqref{defc4}
and
varying it with respect to $f(r)$, $h(r)$, $A_t(r)$, and $A_r(r)$,
we obtain the Euler-Lagrange equations for each of them,
respectively.
Our approach is different from the standard one 
that first one derives the covariant equations of motion by varying the covariant action
and then substitute the ansatz for the metric and vector field
in the static and spherically symmetric spacetime \eqref{ansatz},
and has been employed 
for the investigation of the black hole solutions 
in modified theories of gravitation
\cite{Heisenberg:2017xda,Heisenberg:2017hwb,Kase:2018owh,dhostbh2,dhostbh3,dhostbh4,dhostbh5}.
In Appendix \ref{app_c},
we demonstrate that 
our approach correctly reproduces the Reissner-Nordstr\"{o}m solution
in the simplest Einstein-Maxwell theory.

We will explicitly specify the metric functions $f(r)$ and $h(r)$
to be those of the Schwarzschild, 
Schwarzschild-de Sitter/ anti-de Sitter,
Reissner-Nordstr\"{o}m-type,
and
Reissner-Nordstr\"{o}m de Sitter/ anti-de Sitter-type solutions,
respectively, given by
\begin{itemize}

\item
Schwarzschild solution
\be
\label{sch}
f(r)
=
h(r)
=1-\frac{2M}{r},
\ee
where $M$ is the constant mass parameter of the black hole
measured at the spatial infinity.

\item
Schwarzschild-de Sitter/ anti-de Sitter solution
\be
\label{schds}
f(r)=h(r)
=
1
-\frac{2M}{r}
-\frac{\Lambda}{3}r^2,
\ee
where 
$M$ and $\Lambda$
are constant mass parameter of the black hole
and the effective cosmological constant,
respectively.

\item
Reissner-Nordstr\"{o}m-type solution
\be
\label{rn}
f(r)=h(r)=1-\frac{2M}{r}+\frac{{\cal Q}^2}{r^2},
\ee
where the constant parameter ${\cal Q}$
is related to the electric charge $Q$ (see Eq. \eqref{coulomb}),
depending on the class of the theories.
For example,
in the Einstein-Maxwell theory 
(see Eq. \eqref{rn_em}),
\be
\Q=\frac{Q}{\sqrt{2}M_p},
\ee
where $M_p$ represents the reduced Planck mass \eqref{em}.

\item
Reissner-Nordstr\"{o}m-de Sitter/ anti-de Sitter-type solution
\be
\label{rnds}
f(r)
=h(r)
=1
-\frac{2M}{r}
-\frac{\Lambda}{3}r^2
+\frac{{\cal Q}^2}{r^2},
\ee
where again the constant parameter ${\cal Q}$
is related to the electric charge $Q$.

\end{itemize}

Next, 
along Refs. \cite{Chagoya:2016aar,Minamitsuji:2016ydr,Heisenberg:2017xda,Heisenberg:2017hwb,Minamitsuji:2017aan},
we will consider the vector field with the constant spacetime norm 
\be
\label{y0}
{\cal Y}={\cal Y}_0:={\rm const},
\ee
which yields
\be
\label{a1}
A_r(r)=\pm 
\sqrt{
\frac{\Y_0+A_t(r)^2 /f(r)}
        {h(r)}}.
\ee
Assuming that $A_t(r)$ is regular at the black hole event horizon
as one approaches the event horizon $r=r_g$
characterized by 
\begin{subequations}
\be
&&f(r\to r_g)\to 0,
\\
&&
 h(r\to r_g)\to 0,
\\
&&
\frac{f(r\to r_g)}{h(r\to r_g)}\to {\rm constant},
\ee
\end{subequations}
and hence
\be
A_\mu dx^\mu
\approx 
A_t(r)
\left(
dt\pm dr_\ast
\right)
=A_t(r)
\times
\begin{cases}
    du \\
    dv,
  \end{cases}
\ee
where $dr_\ast:= dr/\sqrt{fh}$
represents the tortoise coordinate,
and $u:=t+r_\ast$ and $v:=t-r_\ast$
are the null coordinates
regular at the future and past event horizons,
respectively.
Hence,
the vector field is regular
at either the future or past event horizon.
For instance, 
$r_g=2M$ in the case of the Schwarzschild solution \eqref{sch}.
The similar discussion can also be applied to
the cosmological horizon, if exists \cite{Minamitsuji:2016ydr}.

More specifically, we consider the following form of $A_t$;

\begin{itemize}

\item
constant $A_t$;
\be
\label{a0q}
A_t(r)=q,
\ee
where $q$ is constant.
Since $F_{tr}=-\partial_r A_t=0$, 
the black hole solutions
do not possess the electric charge.
As argued in Appendix \ref{app_b},
the solutions obtained in this subsection 
can be mapped to
those
in the quadratic-order DHOST theories with the scalar field $\phi$,
via the substitution $A_\mu\to \partial_\mu\phi$
and the integration.

\item
Coulomb form of $A_t$;
\be
\label{coulomb}
A_t(r)
=q+\frac{Q}{r},
\ee
where $Q$ is constant.
Since $F_{tr}=-\partial_r A_t=Q/r^2$, 
the constant $Q$ corresponds to the electric charge
of the vector field.

\end{itemize}

Having the above assumptions for the metric and vector field,
first we substitute the metric functions 
(either \eqref{sch}, \eqref{schds}, \eqref{rn}, or \eqref{rnds}),
the temporal component of the vector field
(either \eqref{a0q} or \eqref{coulomb}),
and 
the radial component of the vector field \eqref{a1}
into each component of the Euler-Lagrange equations,
and 
then to reduce them to a set of the algebraic equations
(see e.g., Refs.
\cite{Chagoya:2016aar,Minamitsuji:2016ydr,Heisenberg:2017xda,Heisenberg:2017hwb,Minamitsuji:2017aan,Babichev:2016kdt,dhostbh2,dhostbh3,dhostbh4}).
In the limit of $r\gg r_g$,
we expand each component of the Euler-Lagrange equations 
in terms of $1/r$.
At each order of the $1/r$ expansion,
we require
that 
the coefficient vanishes,
and  
obtain the condition among
$f_0(\Y)$, $f_2(\Y)$, $\alpha_i (\Y)$
($i=1,2,\cdots, 7,8$) 
and their first-order derivatives $\Y$
evaluated at $\Y=\Y_0$.
We then go to the next order of the $1/r$ expansion
and impose the similar conditions.

We repeat this manipulation
until all the coefficients of the $1/r$ expansion 
of the Euler-Lagrange equations automatically vanish.
When
the Euler-Lagrange equations are satisfied 
at all orders of the $1/r$ expansion, 
the series expansion of the Euler-Lagrange equations 
covers
the entire domain
from the spatial infinity toward the black hole event horizon $r_g<r<\infty$
\cite{Babichev:2016kdt,dhostbh2,dhostbh3,dhostbh4}.
We emphasize that 
that
the conditions for the existence of the black hole solutions
are independent of the choice of the reference point
with respect to which the Euler-Lagrange equatios are expanded,
since in the end the series expansion covers the entire domain
after all the conditions for the existence of the solutions
are determined.
Thus, 
the same tuning relations among
$f_0(\Y)$, $f_2(\Y)$, $\alpha_i (\Y)$
($i=1,2,\cdots, 7,8$) 
and their first-order derivatives $\Y$
evaluated at $\Y=\Y_0$
can be obtained,
even if we expand the Euler-Lagrange equations
with respect to the black hole event horizon $r=r_g$.
Moreover,
if there exists the cosmological horizon
in the case of the asymptotically de Sitter solution, 
the same conditions can be obtained, 
when the Euler-Lagrange equations are
expanded with respect to the cosmological horizon.
We note that 
the solutions of Eqs. \eqref{sch} and \eqref{schds}
correspond to the special cases
of the solutions of  Eqs. \eqref{rn} and \eqref{rnds}
with ${\cal Q}=0$,
which can be obtained 
when the tuning relations
among $f_0(\Y)$, $f_2(\Y)$, $\alpha_i (\Y)$
($i=1,2,\cdots, 7,8$)
evaluated at $\Y=\Y_0$
hold even for $Q\neq 0$ \cite{Chagoya:2016aar}.

We will classify our exact black hole solutions
into the two cases in terms of the spacetime norm of the vector field;

\begin{itemize}

\item
Case-1-$N$ ($N=1,2$); $\Y_0=-q^2$.

The subcases of $N=1$ and $2$
correspond to the cases with $A_t(r)$
given by Eqs. \eqref{a0q} and \eqref{coulomb}, respectively.
``$\Lambda$'' is also attached
in the case that the solution is asymptotically de Sitter or anti-de Sitter.

\item
Case-2-$N$ ($N=1,2$); $\Y_0\neq  -q^2$.

The subcases of $N=1$ and $2$
correspond to the cases with $A_t(r)$
given by Eqs. \eqref{a0q} and \eqref{coulomb}, respectively.
``$\Lambda$'' is also attached
in the case that the solution is asymptotically de Sitter or anti-de Sitter.

\end{itemize}

Finally, 
we check the compatibility of these conditions 
with the degeneracy conditions of the quadratic-order extended vector-tensor theories.
As we will see later, 
in the cases both for the stealth Schwarzschild
and Schwarzschild-de Sitter/ anti-de Sitter solutions,  
in the most cases
our conditions are compatible with the degeneracy conditions, 
and certainly recover the previous results
in the case of the generalized Proca theories.

\subsection{The construction of this paper}
\label{sec1d}

The construction of this paper is as follows;
in Sec. \ref{sec2} and Sec. \ref{sec3},
we obtain the conditions for
the existence of the stealth Schwarzschild 
and Schwarzschild de Sitter/ anti-de Sitter solutions
in the quadratic-order extended vector-tensor theories,
respectively.
In Sec. \ref{sec4},
we check the compatibility of  these conditions with the degeneracy conditions.
In Sec. \ref{sec5},
we discuss the limit of our results 
to the case of the generalized Proca theories.
Sec. \ref{sec6}
is devoted to giving the summary and conclusions.

In Appendix \ref{app_a},
we review the Hamiltonian analysis 
of the theory \eqref{toy}
and how the degeneracy condition removes the Ostrogradsky instabilities.
In Appendix \ref{app_b},
we review the conditions for the existence of 
the steath Schwarzschild and Schwarzschild de Sitter/ anti-de Sitter solutions
in the quadratic-order DHOST theories,
and
the check whether they coincide with the scalar-tensor limits
of the conditions in the extened quadratic-order vector-tensor theories.
In Appendix \ref{app_c},
we illustrate 
that our approach reproduces the Reissner-Nordstr\"{o}m  solution
in the Einstein-Maxwell theory.

\section{The Schwarzschild solutions}
\label{sec2}

In this section, 
we consider the case of the Schwarzschild solution given by Eq. \eqref{sch}.
The black hole event horizon is located at $r=r_g:=2M$.
All the Schwarzschild solutions discussed below
are of stealth type,
in the sense that the mass parameter $M$
is independent of 
the model parameters in 
the coupling functions $f_0(\Y)$, $f_2(\Y)$, 
and $\alpha_i (\Y)$ ($i=1,2,\cdots, 7,8$) \cite{Chagoya:2016aar}.

\subsection{The stealth Schwarzschild solution}
\label{sec2a}

First, we consider the case
that $A_t$ is given by Eq. \eqref{a0q}.
As noted in the subsection \ref{sec1c},
there are the two branches of
the stealth Schwarzschild solutions
given by

\begin{itemize}

\item{Case 1-1}
\begin{subequations}
\label{case11}
\be
&&
f_0=f_{0,\Y}=0,
\\
&&
\alpha_2
=-\alpha_1,
\\
&&
\alpha_{2,\Y}
=-\alpha_{1,\Y}.
\ee
\end{subequations}

\item{Case 2-1}
\begin{subequations}
\label{case21}
\be
&&
f_0=f_{0,\Y}=0,
\\
&&
\alpha_1=
\alpha_2=0,
\\
&&
\alpha_{2,\Y}
=-\alpha_{1,\Y},
\\
&&
\alpha_3
=-2\alpha_{1,\Y}.
\ee
\end{subequations}

\end{itemize}
$f_{0,\Y}$, $f_{2,\Y}$, and $\alpha_{i,\Y}$ ($i=1,2,\cdots,7,8$)
denote the derivatives of 
$f_0(\Y)$, $f_2(\Y)$, and $\alpha_i (\Y)$,
respectively.

It has been shown that Class A
of the quadratic-order extended vector-tensor theories
can be mapped from the quartic-order generalized Proca theories
via the vector disformal transformation \cite{Kimura:2016rzw}. 
In Ref.~\cite{Minamitsuji:2020jvf},
the disformal transformation of the static and stationary black hole solutions
in the vector-tensor theories
has been discussed.
As shown in Ref.~\cite{Minamitsuji:2020jvf},
via the vector disformal transformation
\be
{\tilde g}_{\mu\nu}
=g_{\mu\nu}+\Q\left(\Y\right) A_\mu A_\nu,
\ee
the stealth Schwarzschild solution
with 
the mass $M$ (see Eq. \eqref{sch}),
$A_t=q$  (see Eq. \eqref{a0q}),
and the constant norm $\Y=-q^2$
in a class of the generalized Proca theory
is disformally mapped 
to the stealth Schwarzschild solution 
with the rescaled mass 
\be
\label{resm}
{\tilde M}=\frac{M}{1-\Q(-q^2) q^2},
\ee
where we assume $1-\Q(-q^2)q^2>0$,
which in part explains why 
the stealth Schwarzschild solution
also exists in the other 
(disformally related) quadratic-order extended vector-tensor theories
satisfying Eq. \eqref{case11}.
It is also interesting to note that 
as the special case of Eq. \eqref{resm}
the Minkowski solution 
with $M=0$ and $q\neq 0$
is also mapped to the Minkowski solution
with $\tilde{M}=0$ and $q\neq 0$.
Thus, the disformal transformation
does not modify the vacuum structure
with the nonzero vector field.

\subsection{The charged stealth Schwarzschild solution}
\label{sec2b}

Second, we consider the case given by Eq. \eqref{coulomb}.
Although the black holes are electrically charged,
its contribution does not appear in the spacetime metric \eqref{sch},
and hence we call the solution obtained in this subsection
the {\it charged stealth} Schwarzschild solution,
which was originally obtained 
in the context of the generalized Proca theories
in Refs. \cite{Chagoya:2016aar,Minamitsuji:2016ydr}.

We also obtain the two branches of
the charged stealth Schwarzschild solutions given by
\begin{itemize}

\item{Case 1-2}
\begin{subequations}
\label{case12}
\be
\label{case12a}
f_0
&=&
f_{0,\Y}=0,
\\
\label{case12b}
 \alpha_2
&=&
-\alpha_1,
\\
\label{case12c}
\alpha_{2,Y}
&=&
-\alpha_{1,Y},
\\
\label{case12d}
\alpha_6
&=&
3\alpha_1
+\frac{q^2}{4}
\left[
 2\alpha_3 
-\alpha_4
+\alpha_7
\right.
\nonumber\\
&&
\left.
+
6\alpha_{1,\Y}
-2\alpha_{6,\Y}
+q^2
\left(
 \alpha_{4,\Y}
+ \alpha_{7,\Y}
+\alpha_{8,\Y}
\right)
\right],
\nonumber
\\
&&
\\
\label{case12e}
\alpha_{8}
&=&
\frac{1}{2}
\left[
 2\alpha_3 
-3\alpha_4
-\alpha_7
+6\alpha_{1,\Y}
-2 \alpha_{6,\Y}
\right.
\nonumber\\
&&
\left.
+q^2
\left(
\alpha_{4,\Y}
+
 \alpha_{7,\Y}
+
\alpha_{8,\Y}
\right)
\right],
\ee
\end{subequations}
$\alpha_1$, $\alpha_4$, $\alpha_6$, $\alpha_7$, and $\alpha_8$
satisfy the relation
\be
\label{tuning1}
-
6\alpha_1
+
2\alpha_6
+\Y
\left(
\alpha_4
+\alpha_7
+\alpha_8
\right)
=0.
\ee

\item{Case 2-2}
\begin{subequations}
\label{case22}
\be
\label{case22a}
f_0
&=&
f_{0,\Y}=0,
\\
\label{case22b}
\alpha_1
&=&
\alpha_2
=0,
\\
\label{case22c}
\alpha_{2,\Y}
&=&
-\alpha_{1,\Y},
\\
\label{case22d}
\alpha_3
&=&
-2\alpha_{1,\Y},
\\
\label{case22e}
\alpha_6
&=&
\frac{\Y_0}{4}
\left[
\alpha_4
-\alpha_7
-2\alpha_{1,Y}
+2 \alpha_{6,\Y}
\right.
\nonumber\\
&&
\left.
+\Y_0
\left(
\alpha_{4,\Y}
+\alpha_{7,\Y}
+\alpha_{8,\Y}
\right)
\right],
\\
\label{case22f}
\alpha_{8}
&=&
-
\frac{1}{2}
\left[
 3\alpha_4
+\alpha_7
-2\alpha_{1,\Y}
+2 \alpha_{6,\Y}
\right.
\nonumber\\
&&
\left.
+\Y_0
\left(
\alpha_{4,\Y}
+\alpha_{7,\Y}
+\alpha_{8,\Y}
\right)
\right],
\ee
\end{subequations}
$\alpha_4$, $\alpha_6$, $\alpha_7$, and $\alpha_8$
satisfy the relation
\be
\label{f42}
2\alpha_6
+\Y_0
\left(
\alpha_4
+\alpha_7
+\alpha_8
\right)
=0.
\ee

\end{itemize}

In Case 1-2,
when the tuning relation \eqref{case12d}
is not satisfied,
the spacetime metric effectively
reduces to the form of the Reissner-Nordstr\"{o}m-type solution
given by Eq. \eqref{rn}
with
\be
\label{calq}
{\cal Q}
= 
{\cal Q}_1
&:=&
\frac{Q}
{
\sqrt{2
\left[
 8f_2
+6q^2\alpha_1
+q^4
\left(
  \alpha_3
+4\alpha_{1,\Y}
\right)
\right]}}
\nonumber\\
&\times&
\Big[
 12\alpha_1
+2 q^2\alpha_3
-q^2\alpha_4
-4\alpha_6
\nonumber\\
&+&
q^2\alpha_7
+6q^2\alpha_{1\Y}
-2q^2\alpha_{6,\Y}
\nonumber\\
&+&
q^4
\left(
\alpha_{4,\Y}
+ \alpha_{7,\Y}
+\alpha_{8,\Y}
\right)
\Big]^{\frac{1}{2}}.
\ee
While Eqs. \eqref{case12a}-\eqref{case12c}
remain the same,
as the consequence of ${\cal Q}_1\neq 0$, 
Eq.~\eqref{case12e} is modified.

In Case 2-2,
when the tuning relation \eqref{case22e}
is not satisfied,
the spacetime metric effectively
reduces to the form of the Reissner-Nordstr\"{o}m-type solution
\eqref{rn}
with
\be
\label{calq2}
{\cal Q}_2
=
{\cal Q}
&:=&
\frac{Q}
       {2\sqrt{4f_2+\Y_0^2\alpha_{1,\Y}}}
\Big[ 
  \Y_0 \alpha_4
-4\alpha_6
\nonumber\\
&+&
\Y_0
\left(
-\alpha_{7}
-2\alpha_{1,\Y}
+
2\alpha_{6,\Y}
\right.
\nonumber\\
&&
\left.
+\Y_0 
\left(
 \alpha_{4,\Y}
+\alpha_{7,\Y}
+\alpha_{8,\Y}
\right)
\right)
\Big]^{\frac{1}{2}}.
\ee
While
Eqs. \eqref{case22a}-\eqref{case22d}
remain the same,
as the consequence of ${\cal Q}_2\neq 0$, 
Eq. \eqref{case22f} is modified.

\section{The Schwarzschild-de Sitter/ anti-de Sitter solutions}
\label{sec3}

Second, we consider the Schwarzschild-de Sitter / anti-de Sitter solutions \eqref{schds}.
The black hole event horizon is located at $r=r_g$,
which corresponds to the smallest positive root
of the equation $f=h=0$.
We will {\it not} call
the Schwarzschild de Sitter/ anti-de Sitter solutions
obtained obtined in this section
{\it stealth} solutions,
since the effective cosmological constant $\Lambda$
depends on the coupling functions 
$f_0(\Y)$, $f_2(\Y)$, and $\alpha_i(\Y)$
($i=1,2,\cdots, 7,8$).

\subsection{The Schwarzschild-de Sitter/ anti-de Sitter  solution}
\label{sec3a}

First, we consider the case Eq. \eqref{a0q}.
As noted in the subsection \ref{sec1c},
the solutions discussed in this subsection 
can be mapped to
those in the quadratic-order DHOST theories with the scalar field
(see Appendix \ref{app_b}).

We find the two branches of
the Schwarzschild-de Sitter/ anti- de Sitter solutions,
given by Eq. \eqref{schds};
\begin{itemize}

\item{Case 1-1-$\Lambda$}
\begin{subequations}
\label{case13}
\be
&&
\label{case13a}
f_0=
-2\Lambda
\left(
  f_2
+q^2\alpha_1
\right),
\\
&&
\label{case13b}
f_{0,\Y}=
\Lambda
\left[
\alpha_1
+
\frac{3}{2}q^2
\alpha_3
-
4f_{2,\Y}
+
2q^2\alpha_{1,\Y}
\right],
\\
&&
\label{case13c}
\alpha_2
=-\alpha_1,
\\
&&
\label{case13d}
\alpha_{2,\Y}
=-\alpha_{1,\Y}.
\ee
\end{subequations}

\item{Case 2-1-$\Lambda$}
\begin{subequations}
\label{case23}
\be
f_0
&=&-2\Lambda f_2,
\\
f_{0,\Y}
&=&
\Lambda
\left(
-4f_{2,\Y}
+\Y_0 \alpha_{1,\Y}
\right),
\\
\alpha_1
&=&
\alpha_2
=0,
\\
\alpha_{2,\Y}
&=&
-\alpha_{1,\Y},
\\
\alpha_3
&=&
-2\alpha_{1,\Y}.
\ee
\end{subequations}

\end{itemize}

\subsection{The charged Schwarzschild-de Sitter/ anti-de Sitter solution}
\label{sec3b}

Second, we consider the case \eqref{coulomb},
where
we find the two branches of the charged Schwarzschild-de Sitter/ anti- de Sitter solutions, given by
\begin{itemize}

\item{Case 1-2-$\Lambda$}
\begin{subequations}
\label{case14}
\begin{eqnarray}
\label{case14a}
f_{0}
&=&
-2\Lambda
\left(
f_2
+
q^2\alpha_1
\right),
\\
\label{case14b}
f_{0,\Y}
&=&
\frac{\Lambda}{2}
\left[
  2\alpha_1
-8 f_{2,\Y}
+q^2
\left(
  3\alpha_3
+4\alpha_{1,\Y}
\right)
\right],
\qquad 
\\
\label{case14c}
\alpha_2
&=&
-\alpha_1,
\\
\label{case14d}
\alpha_{2,\Y}
&=&
-\alpha_{1,\Y},
\\
\label{case14e}
\alpha_6
&=&
3\alpha_1
+\frac{q^2}{4}
\left[
 2\alpha_3 
-\alpha_4
+\alpha_7
+6\alpha_{1,\Y}
-2\alpha_{6,\Y}
\right.
\nonumber\\
&&
\left.
+q^2
\left( 
\alpha_{4,\Y}
+\alpha_{7,\Y}
+\alpha_{8,\Y}
\right)
\right],
\\
\label{case14f}
\alpha_{8}
&=&
\frac{1}{2}
\Big[ 
 2\alpha_3 
-3\alpha_4
-\alpha_7
+6\alpha_{1,\Y}
-2 \alpha_{6,\Y}
\nonumber\\
&+&
q^2 
\Big(
\alpha_{4,\Y}
+
\alpha_{7,\Y}
+
\alpha_{8,\Y}
\Big)
\Big],
\end{eqnarray}
\end{subequations}
which satisfies Eq. \eqref{tuning1}.

\item{Case 2-2-$\Lambda$}
\begin{subequations}
\label{case24}
\be
\label{case24a}
f_0
&=&
-2\Lambda f_2,
\\
\label{case24b}
f_{0,\Y}
&=&
\Lambda
\left(
-4 f_{2,\Y}
+\Y_0 \alpha_{1,\Y}
\right),
\\ 
\label{case24c}
\alpha_1
&=&
0,
\\
\label{case24d}
\alpha_2
&=&0,
\\
\label{case24e}
\alpha_{2,\Y}
&=&
-\alpha_{1,\Y},
\\
\label{case24f}
\alpha_3
&=&
-2\alpha_{1,\Y},
\\
\label{case24g}
\alpha_6
&=&
\frac{\Y_0}{4}
\left[
\alpha_4
-\alpha_7
-2\alpha_{1,\Y}
+2 \alpha_{6,\Y}
\right.
\nonumber\\
&&
\left.
+\Y_0
\left(
 \alpha_{4,\Y}
+\alpha_{7,\Y}
+ \alpha_{8,\Y}
\right)
\right],
\\
\label{case24h}
\alpha_{8}
&=&
-
\frac{1}{2}
\left[
 3\alpha_4
+\alpha_7
-2\alpha_{1,\Y}
+2 \alpha_{6,\Y}
\right.
\nonumber\\
&&
\left.
+\Y_0
\left(
 \alpha_{4,\Y}
+\alpha_{7,\Y}
+ \alpha_{8,\Y}
\right)
\right],
\ee
\end{subequations}
which satisfy Eq. \eqref{f42}.

\end{itemize}

In Case 1-2-$\Lambda$,
when the tuning relation \eqref{case14e}
is not satisfied,
the spacetime metric 
effectively
reduces to the form of the Reissner-Nordstr\"{o}m-de Sitter/anti-de Sitter
type solution given by \eqref{rnds}
with ${\cal Q}={\cal Q}_1$ as in Eq. \eqref{calq}
While
$\Lambda$ satisfies Eqs.~\eqref{case14a}-\eqref{case14b}
and Eqs.~\eqref{case14a}-\eqref{case14d}
remain the same,
as the consequence of ${\cal Q}_1\neq 0$, 
Eq.~\eqref{case14f} is modified.

In Case 2-2-$\Lambda$,
when the tuning relation \eqref{case24g} is not satisfied, 
the spacetime metric effectively reduces to the form of the Reissner-Nordstr\"{o}m-de Sitter/anti-de Sitter
solution Eq. \eqref{rnds}
with ${\cal Q}={\cal Q}_2$ as in Eq. \eqref{calq2}.
While
$\Lambda$ satisfies Eqs.~\eqref{case24a}-\eqref{case24b}
and Eqs.~\eqref{case24a}-\eqref{case24f}
remain the same,
as the consequence of ${\cal Q}_2\neq 0$, 
Eq.~\eqref{case24h} is modified.

In the above examples,
$\Lambda$ is determined not by the bare value of the cosmological
constant but by the parameters in the coupling functions.
Such a feature would be crucial
for the realization of the self-tuning of the cosmological constant,
which was originally suggeted in the context of the Horndeski theory 
\cite{Babichev:2013cya}.

\section{The compatibility with the degeneracy conditions}
\label{sec4}

\subsection{The degeneracy conditions}
\label{sec4a}

The degeneracy conditions
for the quadratic-order extended vector-tensor theory \eqref{action} with Eq. \eqref{defc4}
given in Ref. \cite{Kimura:2016rzw}
were summarized in subsection \ref{sec1b}.
There are the following subclasses in Class A;

\begin{itemize}

\item
Class A1
\begin{subequations}
\label{class_a1}
\be
\label{class_a1a}
\alpha_1
&=&-\alpha_2
=
\frac{f_2}{\Y},
\\
\label{class_a1b}
\alpha_3
&=&
\frac{2(f_2-2f_{2,\Y} \Y)}{\Y^2},
\ee
\end{subequations}
which requires $\Y\neq 0$.

\item
Class A2
\begin{subequations}
\label{class_a2}
\be
\alpha_1&=&-\alpha_2
=\frac{f_2}{\Y},
\\
\alpha_4
&=&\frac{6f_2+\Y\beta}{\Y^2}
-\alpha_8,
\ee
\end{subequations}
where 
\be
\beta:=-2\alpha_6-\alpha_7\Y,
\ee
which requires $\Y\neq 0$.

\item
Class A3
\begin{subequations}
\label{class_a3}
\be
\label{class_a3a}
\alpha_1
&=&
-\alpha_2
=
-\frac{(\alpha_4+\alpha_8)\Y-\beta}
         {2},
\\
\alpha_3
&=&
\frac{1}
       {4f_2}
\left(
(\alpha_4+\alpha_8)\Y -\beta
\right)
\nonumber\\
&\times&
\left[
\left(
-2\beta
+8f_{2,\Y}
+\alpha_8 \Y
\right)
\right.
\nonumber\\
&&
\left.
-2
(2\alpha_4+\alpha_8)
f_2
\right],
\ee
\end{subequations}
which requires $f_2\neq 0$.

\item
Class A4
\begin{subequations}
\label{class_a4}
\be
\alpha_1
&=&
-\alpha_2,
\\
\alpha_4
&=&
\frac{3\left(2\alpha_2+4f_{2,\Y}\right)}
          {8(f_2+\alpha_2 \Y)}
-\alpha_3
-\alpha_5\Y,
\\
\alpha_8
&=&
\frac{-W_2\pm \sqrt{W_2^2-2W_1W_3}}
        {2W_1},
\ee
\end{subequations}
where for the definition of $W_k$ ($k=1,2,3$)
see (3.28)-(3.30) of  Ref. \cite{Kimura:2016rzw}.
From (3.29a) in Ref. \cite{Kimura:2016rzw},
$W_1\neq 0$ requires
that 
$\Y\neq 0$ and $f_2+\alpha_2 \Y\neq 0$.

\end{itemize}


In Class A2,
the second relation in Eq.~\eqref{class_a2}
can be rewritten as 
\be
-6\alpha_1
+2\alpha_6
+
\left(
 \alpha_4
+\alpha_7
+\alpha_8
\right)
\Y
=0.
\ee
In the case that
the general theory \eqref{action}
with Eq.~\eqref{defc4}
admits a solution with $\alpha_1=\alpha_2=0$.

In Class A3,
the relation \eqref{class_a3a}
can be rewritten as 
\be
2
\left(
\alpha_1+\alpha_6
\right)
+\Y 
\left(
  \alpha_4
+\alpha_7
+\alpha_8
\right)
=0.
\ee
In the case that
the general theory \eqref{action}
with Eq.~\eqref{defc4}
admits a solution with $\alpha_1=\alpha_2=0$,
on this background
Eq. \eqref{class_a3a}
suggests
\be
2\alpha_6
+\Y 
\left(
  \alpha_4
+\alpha_7
+\alpha_8
\right)
=0.
\ee

\subsection{The compatibility with the degeneracy conditions}
\label{sec4b}

The compatibility of the conditions for the existence of the solutions 
with the degeneracy conditions 
are summarized in Table \ref{table1}.
\begin{table}
\begin{tabular}{c| c c c c}
Class \textbackslash Case
 & 1-1(-$\Lambda$) & 1-2(-$\Lambda$) & 2-1(-$\Lambda$) & 2-2(-$\Lambda$)  
\\ 
 \hline
A1     & $\checkmark$
         & $\checkmark$
         & $\checkmark$
         & $\checkmark$
\\
A2     & $\checkmark$
         & $\checkmark$
         & $\checkmark$
         & $\checkmark$
\\
A3     & $\checkmark$\footnotemark[1]
         & $\checkmark$\footnotemark[1]
         & $\checkmark$\footnotemark[1]
         & $\checkmark$\footnotemark[1]
 \\
A4     & $\checkmark$\footnotemark[2]
         & $\checkmark$\footnotemark[2]
         & $\checkmark$\footnotemark[2]
         & $\checkmark$\footnotemark[2]
\\  
\hline
B       & $\times$ & $\times$ &$\times$ & $\times$  
\\   
\end{tabular}
\caption{
The compatibility with the degeneracy conditions.
``Class'' represents 
the classes of the quadratic-order extended vector-tensor theories
(see Secs. \ref{sec1b} and \ref{sec4a}),
which are the solutions of the degeneracy conditions, 
while 
``Case'' represents 
the conditions for obtaining the exact black hole solutions
(see Secs. \ref{sec2} and \ref{sec3}).
Footnotes ``a'' and ``b''
represent 
the conditions 
under which the compatibility 
of the conditions for the existence of the exact black hole solutions
with the degeneracy conditions hold.
}
\label{table1}
\footnotetext[1]{
$f_2\neq 0$ evaluated at $\Y=\Y_0$.}
\footnotetext[4]{
$\Y\neq 0$ and 
$f_2+\alpha_2\Y\neq 0$ 
evaluated at $\Y=\Y_0$.}
\end{table}
The columns with $\checkmark$
correspond to the cases which are consistent 
with the degeneracy conditions
under the assumptions shown in the footnotes,
while 
those with $\times$
correspond to the cases which cannot be consistent.

All the solutions of 
Case 1-1, 
Case 1-2, 
Case 2-1,
Case 2-2,
Case 1-1-$\Lambda$, 
Case 1-2-$\Lambda$, 
Case 2-1-$\Lambda$, 
and 
Case 2-2-$\Lambda$
(Eqs. \eqref{case11}, \eqref{case12}, \eqref{case21}, \eqref{case22},
\eqref{case13}, \eqref{case14}, \eqref{case23}, and \eqref{case24},
respectivelty)
are
not compatible with Class B,
since $\alpha_1+\alpha_2\neq 0$ at $\Y=\Y_0$.
On the other hand,
all the solutions
are compatible with Class A.

\subsection{The correspondence to the scalar-tensor theories}

The vector-tensor theories \eqref{action} with Eq. \eqref{defc4}
reduce to 
the shift-symmetric scalar-tensor theories \eqref{action2} with Eq. \eqref{c_sca},
in the limit of 
$A_\mu\to \partial_\mu \phi$,
$\Y\to \X$ (see Eq. \eqref{defx} for the definition of the canonical kinetic term of the scalar field), 
and 
$\alpha_i= 0$ ($i=6, 7, 8$).
The vector field with the constant $A_t$,
Eq. \eqref{ansatzb}
with Eq. \eqref{a0q},
then reduces to 
\be
\phi= qt +\psi(r),
\qquad
\psi(r)
:=
\int dr A_r(r).
\ee
We confirm that 
the conditions for the existence of 
the black hole solutions with the vanishing electric field,
Case 1-1, 
Case 2-1,
Case 1-1-$\Lambda$, 
and 
Case 2-1-$\Lambda$
(Eqs. \eqref{case11}, \eqref{case21}, \eqref{case13}, and \eqref{case23}),
reduce
to 
those of 
Case 1,
Case 2,
Case 1-$\Lambda$,
and 
Case 2-$\Lambda$
obtained in Refs. \cite{dhostbh2} 
(see also Refs. \cite{dhostbh1,Takahashi:2020hso}),
which will be reviewed in Appendix \ref{app_a}.

\subsection{On the black hole perturbations and the strong coupling problem}

As shown in the previous subsections,
the stealth Schwarzschild,
the charged stealth Schwarzschild,
the Schwarzschild de Sitter/ anti-de Sitter,
and 
the charged Schwarzschild de Sitter/ anti-de Sitter
solutions
exist in
the Class A theories.
The limit of them 
the generalized Proca theories 
will be discussed in Sec.~\ref{sec5}.
Although at the background level
the spacetime geometry remains that of 
the Schwarzschild or Schwarzschild de Sitter/ anti-de Sitter solution,
at the level of the linear perturbations
the vector field degrees of freedom would propagate,
which results in the existence of the extra polarization modes of gravitational waves.
The existence of the extra polarization modes
would be tested by 
the current and future observations of gravitational waves
\cite{Chamberlin:2011ev,Hagihara:2019ihn,Barack:2018yly}.

Regarding the perturbation analysis and the stability,
the possible issues are raised
by referring to the recent studies on the stability of black hole solutions 
in the quadratic-order DHOST theories
\cite{Takahashi:2019oxz,deRham:2019gha,Charmousis:2019fre}.
In the case of the quadratic-order DHOST theories, 
it has been shown that 
the even-parity perturbations on
the stealth Schwarzschild 
and Schwarzschild-de Sitter solutions
suffer from the strong coupling problem \cite{deRham:2019gha},
which spoils the predictability of the perturbation theory.
The similar strong coupling problem may exist for 
the black hole solutions
in the quadratic-order extended vector-tensor theories
reported in this paper.

In the case of the quadratic-order DHOST theories, 
a resolution to this problem
by the controllable violation of the degeneracy conditions
was argued in Ref. \cite{Motohashi:2019ymr}.
It has been shown that 
the effective field theories
about the stealth Minkowski and de Sitter solutions
predict the universal dispersion relation $\omega/M=c \left(k/M\right)^2$,
where 
$c$ is the dimensionless constant 
and 
$M$ represents the scale charactering the background scalar field,
and
are weakly coupled at the energy scales up to $M$ for $c={\cal O} (1)$.
On the other hand, 
in the degenerate theories
where the perturbations are forced to obey
the second-order equations of motion, $c=0$,
indicating the appearance of the strong coupling problem~\cite{Motohashi:2019ymr}.
Thus, the controllable violation of the degeneracy conditions
makes the perturbations weakly coupled.

The similar resolution to the strong coupling problem
may also be able to be applied to the stealth solutions
in the quadratic-order extended vector-tensor theories.
The conditions for the existence of the (charged) stealth Schwarzschild solutions
and the (charged) Schwarzschild-de Sitter/ anti-de Sitter solutions
obtained in this paper
should include the case 
where the degeneracy conditions are controllably violated
within the framework of the general vector-tensor theories \eqref{action} with Eq. \eqref{defc4}.

\section{The special cases in the generalized Proca theories}
\label{sec5}

\subsection{The generalized Proca theories}

In this section, we apply the results in the previous sections
to the quadratic- and quartic-order  generalized Proca theories
\be
\label{gp}
S&=&\int d^4x \sqrt{-g}
\left[
  G_2(\Y)
+G_4  (\Y)R
-\frac{1}{4}F^{\mu\nu}F_{\mu\nu}
\right.
\nonumber\\
&-&
\left.
2G_{4,\Y} (\Y)
\left(
  \left( \nabla^\mu A_\mu\right)^2
-\nabla^\mu A^\nu \nabla_\nu A_\mu
\right)
\right],
\ee
where $G_2$ and $G_4$ are the functions of $\Y$,
which corresponds to the following choice of 
the quadratic-order extended vector-tensor theories
\eqref{action} with Eq.~\eqref{defc4}
\begin{subequations}
\label{gp2}
\begin{eqnarray}
f_0
&=&G_2,
\\
{f}_2
&=&{G}_{4},
\\
{\alpha}_1
&=&
-{\alpha}_2
=
2{G}_{4,{\cal Y}},
\\
{\alpha}_6
&=&
-2{G}_{4,{\cal Y}}-1,
\\
{\alpha}_3
&=&
{\alpha}_4
=
{\alpha}_5
=
{\alpha}_7
=
{\alpha}_8
=0.
\end{eqnarray}
\end{subequations}
Since $W_2=W_3=0$ in Eq. \eqref{class_a4},
the generalized Proca theories \eqref{gp} 
belong to the Class A4 theories.

The conditions
for the stealth Schwarzschild and Schwarzschild-de Sitter/ anti-de Sitter solutions  
obtained in the previous sections reduce as follows:

\begin{itemize}

\item Case 1-1

The conditions \eqref{case11} reduce to
\be
\label{case11gp}
G_2=G_{2,\Y}=0.
\ee

\item Case 2-1

The conditions \eqref{case21} reduce to
\be
\label{case21gp}
G_2=G_{2,\Y}=G_{4,\Y}=G_{4,\Y\Y}=0.
\ee

\item Case 1-2

The conditions \eqref{case12} reduce to
\begin{subequations}
\label{case12gp}
\be
&&
1+8 G_{4,\Y}=0,
\\
&&
G_2
=
G_{2,\Y}
=
G_{4,\Y\Y}=0.
\ee
\end{subequations}

\item Case 1-1-$\Lambda$

The conditions \eqref{case13} reduce to
\begin{subequations}
\label{case13gp}
\be
&&
G_2
+2
\Lambda
\left(G_4+2q^2G_{4,\Y}\right)=0,
\\
&&
G_{2,\Y}
+2\Lambda
\left(
G_{4,\Y}
-2q^2G_{4,\Y\Y}
\right)
=0.
\ee
\end{subequations}

\item Case 2-1-$\Lambda$

The conditions \eqref{case23} reduce to
\begin{subequations}
\label{case23gp}
\be
&&G_2+2\Lambda G_4=0,
\\
&&G_{2,\Y}=G_{4,\Y}=G_{4,\Y\Y}=0.
\ee
\end{subequations}

\item Case 1-2-$\Lambda$

The conditions \eqref{case14} reduce to
\begin{subequations}
\label{case14gp}
\be
&&
G_2
-
2\Lambda
\left(
 G_{4,\Y}
 +2q^2
 G_{4,\Y}
\right)
=0,
\\
&&
G_{2,\Y}
+
2\Lambda
G_{4,\Y}
= 0,
\\
&&
1+8 G_{4,\Y}=0,
\\
&&
G_{4,\Y\Y}=0.
\ee
\end{subequations}

\item Case 2-2 and Case 2-2-$\Lambda$

The conditions \eqref{case22} and \eqref{case24} are not consistent 
in the generalized Proca theories \eqref{gp}.

\end{itemize}

\subsection{The nonminimal coupling to the Einstein tensor}

Furthermore, 
we consider the special case
\begin{subequations}
\be
&&
G_2=-m^2\Y-M_p^2 V_0,
\\
&&
G_4=\frac{M_p^2}{2}-\frac{\beta}{2}\Y,
\ee
\end{subequations}
where $M_p$ and $V_0$
represent the reduced Planck mass and the cosmological constant,
and 
$m$ and $\beta$ are the mass and the coupling constant of the vector field $A_\mu$,
respectively,
which,
up to the total derivative terms,
gives rise to 
the 
generalized Proca theory
with the nonminimal coupling to the Einstein tensor
\cite{Heisenberg:2014rta,Tasinato:2014eka,DeFelice:2016cri}
\begin{eqnarray}
\label{gp3}
S
&=&
\int d^4 x
  \sqrt{-g}
\left[
\frac{M_p^2}{2}
\left(
R-2V_0
\right)
\right.
\nonumber\\
&-&
\left.
\frac{1}{4}
F_{\mu\nu}F^{\mu\nu}
-(m^2g^{\mu\nu}-\beta G^{\mu\nu})
A_\mu A_\nu
\right].
\end{eqnarray}
Hence, $\beta$ corresponds to 
the nonminimal coupling constant
to the Einstein tensor.

\begin{itemize}

\item Case 1-1

The conditions \eqref{case11gp} reduce to
\be
m=V_0=0,
\ee
whcih reproduced
the stealth Schwarzschild solution
with the vanishing electric field
in the generalized Proca theory \eqref{gp3}
discussed in Refs. \cite{Chagoya:2016aar,Minamitsuji:2016ydr}.

\item Case 2-1

The conditions \eqref{case21gp} reduce to
\be
m=\beta=V_0=0,
\ee
which means the Schwarzschild solution 
in general relativity without the cosmological constant
and reproduces the stealth Schwarzschild solution 
discussed in Refs. \cite{Chagoya:2016aar,Minamitsuji:2016ydr}. 
Since the $U(1)$ gauge symmetry is restored,
the solution describes the Schwarzschild solution
with the vanishing electric field in the Einstein-Mawell theory.

\item Case 1-2

The conditions \eqref{case12gp} reduce to
\begin{subequations}
\be
&&m=V_0=0,
\\
&&\beta=\frac{1}{4},
\ee
\end{subequations}
which reproduces the charged stealth Schwarzschild solution 
discussed in Refs. \cite{Chagoya:2016aar,Minamitsuji:2016ydr}.

\item Case 1-1-$\Lambda$

The conditions \eqref{case13gp} reduce to
\begin{subequations}
\be
\Lambda
&=&-\frac{m^2}{\beta},
\\
q
&=&
\pm 
\frac{M_p}
        {\sqrt{2}}
\sqrt{
\frac{1}{\beta}
+\frac{V_0}{m^2}
},
\ee
\end{subequations}
where we have to impose
$1/\beta+V_0/m^2\geq 0$
which reproduces the Schwarzschild-(anti-) de Sitter solution 
discussed in Refs. \cite{Minamitsuji:2016ydr}.

\item Case 2-1-$\Lambda$

The conditions \eqref{case23gp} reduce to
\be
m=\beta=0,
\qquad 
\Lambda=V_0.
\ee

Since the $U(1)$ gauge symmetry is restored,
the solution describes the Schwarzschild-de Sitter/ anti-de Sitter solution
with the vanishing electric field
in the Einstein-Mawell theory with the cosmological constant $V_0$.

\item Case 1-2-$\Lambda$

The conditions \eqref{case14gp} reduce to
\begin{subequations}
\be
\beta&=&\frac{1}{4},
 \\
\Lambda&=&-4m^2,
\\
q
&=&
\pm 
\frac{M_p}
        {\sqrt{2}}
\sqrt{4+\frac{V_0}{m^2}},
\ee
\end{subequations}
where we have to impose $4m^2+V_0\geq 0$
and $m\neq 0$,
which reproduces the charged Schwarzschild-anti-de Sitter solutions
discussed in Ref. \cite{Minamitsuji:2016ydr}.

\end{itemize}

\section{Conclusions}
\label{sec6}

We have investigated the static and spherically symmetric black hole solutions 
in the quadratic-order extended  vector-tensor theories 
without the Ostrogradsky instabilities, 
which include the generalized Proca theories as the particular subclass. 
The theories are given by Eq. \eqref{action} with Eq. \eqref{defc4},
where the free functions of the spacetime norm of the vector field,
$\Y$ defined in Eq, \eqref{norm},
satisfy the degeneracy conditions summarized in Sec \ref{sec4}.
The most general vector-tensor theories \eqref{action} with Eq. \eqref{defc4}
constructed with up to the quadratic-order terms of the first-order covariant 
derivatives of the vector field 
are not free from the Ostrogradsky instabilities,
unless
the certain degeneracy conditions are imposed \cite{Kimura:2016rzw}.

We have started from the most general action of the vector-tensor theories
\eqref{action} with Eq. \eqref{defc4},
and derived the Euler-Lagrange equations for the metric and vector field variables 
in the static and spherically symmetric backgrounds. 
We then substituted the metric functions for 
the Schwarzschild and Schwarzschild-de Sitter/ anti-de Sitter spacetimes 
and the vector field with the constant spacetime norm $\Y=\Y_0$
into the Euler-Lagrange equations.
Under our ansatz \eqref{ansatz} and assumptions,
the vector field which is regular at either the future or past event horizon,
depending on the choice of the branch.

The series expansion analysis of the Euler-Lagrange equations
after the substitutions of the above ansatz 
yielded the conditions for the existence of 
the Schwarzschild and Schwarzschild-de Sitter/ anti-de Sitter solutions
on the free functions of the spacetime norm of the vector field $\Y$,
evaluated at the constant value of $\Y=\Y_0$.
As in the previous cases
on the black hole solutions in the generalized Proca theories
\cite{Chagoya:2016aar,Minamitsuji:2016ydr,Babichev:2017rti,Heisenberg:2017xda,Heisenberg:2017hwb},
we have derived
the conditions for the existence of 
the stealth Schwarzschild,
the charged stealth Schwarzschild,
the Schwarzschild-de Sitter/ anti- de Sitter solutions,
and 
the charged Schwarzschild-de Sitter/ anti- de Sitter solutions,
where
the metric functions do not depend on the electric charge $Q$.
When the tuning relation on the function $\alpha_6$ 
evaluated at $\Y=\Y_0$
is broken,
then the solution 
takes the form of the Reissner-Nordstr\"{o}m-type solution
and
Reissner-Nordstr\"{o}m-de Sitter/ anti- de Sitter-type solution
with the effective charge ${\cal Q}$
(see Eqs. \eqref{calq} and \eqref{calq2}).

Second, we have compared the conditions for the existence of the black hole solutions
of the black hole solutions obtained in the present paper 
with the degeneracy conditions.
We have shown
that 
the conditions for
the existence
the stealth Schwarzschild,
the charged stealth Schwarzschild,
the Schwarzschild-de Sitter/ anti- de Sitter solutions,
and 
the charged Schwarzschild-de Sitter/ anti- de Sitter solutions
are compatible 
with the degeneracy conditions for the Class A theories,
while
they are not compatible with the degeneracy conditions
for the Class B  theories
(see Table \ref{table1}).
We have also recovered the stealth Schwarzschild, charged stealth Schwarzschild, 
and Schwarzschild-de Sitter solution 
in the limit to the  generalized Proca theories.

Although at the background level
the spacetime geometry remains that of 
the Schwarzschild 
and Schwarzschild de Sitter/ anti-de Sitter
solutions,
at the level of the linear perturbations
the extra polarization modes of gravitational waves
would also propagate,
which would be tested by the current and future observations of gravitational waves
\cite{Chamberlin:2011ev,Hagihara:2019ihn,Barack:2018yly}.
As in the case of the static and spherically symmetric 
black hole solutions in the quadratic-order DHOST theories
\cite{deRham:2019gha},
for the solutions satisfying the degeneracy conditions
the strong coupling problem may exist
in the even parity sector of the black hole perturbations, 
which spoils the predictability of the linear perturbation theory.
The strong coupling problem 
may be able to be resolved 
by the controllable violation of the degeneracy conditions
as suggested in Ref. \cite{Motohashi:2019ymr}.
The conditions for the existence of the (charged) stealth Schwarzschild solutions
and the (charged) Schwarzschild-de Sitter/ anti-de Sitter solutions
obtained in this paper
should admit the solutions 
which would not suffer from the strong coupling problem 
within the framework of the theory \eqref{action} with Eq. \eqref{defc4}.

Finally, 
it will be very interesting 
to construct
the black hole solutions with the nonconstant
spacetime norm $\Y=\Y(r)$
and 
the solutions of relativistic stars
in the case of the static and spherically symmetric spacetime.
As the next step. 
it will also be important 
to explore the black hole solutions beyond the spherical symmetry,
e.g., the stationary and axisymmetric black hole solutions.
We hope to come back to these issues in our future publication.

\acknowledgments{
M.M.~was supported by the Portuguese national fund 
through the Funda\c{c}\~{a}o para a Ci\^encia e a Tecnologia
in the scope of the framework contract foreseen
in the numbers 4, 5 and 6 of the article 23,of the Decree-Law 57/2016 
of August 29, changed by Law 57/2017 of July 19,
and the Centro de Astrof\'{\i}sica e Gravita\c c\~ao (CENTRA)
through the Project~No.~UIDB/00099/2020.
}

\appendix 

\section{The Ostrogradsky instabilities and degenerate theories in analytical mechanics}
\label{app_a}

We analyze the analytical toy model \eqref{toy} in the Hamiltonian analysis.
More specifically,
we consider the potential composed of the quadratic-order terms
\begin{eqnarray}
V(x,y)
=\frac{g_1}{2}x^2
+g_2 xy
+\frac{g_3}{2}y^2,
\end{eqnarray}
with $g_1$, $g_2$, and $g_3$ being constants.
The theory equivalent to Eq. \eqref{toy}
can be obtained
by introducing the auxiliary variable $Q$ 
\be
\label{3rd2}
L_{p,2}
&=&
\frac{a_1}{2}\dot{Q}^2
+a_2 \dot{Q} \dot{y}
+\frac{a_3}{2}\dot{y}^2
+\frac{1}{2} \dot{x}^2
-V(x,y)
\nonumber\\
&+&
\lambda
\left(\dot{x}-Q\right).
\ee
We define their conjugate momenta to $x$, $Q$, and $y$,
respectively,
by
\be
&&
P_Q
:= \frac{\partial L_{p,2}}
                 {\partial \dot{Q}}
=a_1\dot{Q}+a_2\dot{y},
\\
&&
P_y:=
\frac{\partial L_{p,2}}
                 {\partial \dot{y}}
= a_3\dot{y}+a_2 \dot{Q},
\\
&&
P_x
:=
\frac{\partial L_{p,2}}
                 {\partial \dot{x}}
= \lambda.
\ee

\subsection{The nondegenerate case}

First, we consider the {\it nondegenerate} case;
\be
\label{dege}
\frac{\partial^2 L_{p,2}}{\partial \dot{Q}^2}
\frac{\partial^2 L_{p,2}}{\partial \dot{y}^2}
-
\left(\frac{\partial^2 L_{p,2}}{\partial \dot{Q}\partial\dot{y}}\right)^2
=
a_1a_3-a_2^2
\neq 0.
\ee
By rewriting $\dot{Q}$ and $\dot{q}$ in terms of $P_Q$ and $P_q$, 
we obtain the Hamiltonian 
\be
\label{ham_ng}
H_p
&:=&
  P_Q\dot{Q}
+P_x\dot{x}
+P_y\dot{y}
-L_{p,2}
\nonumber\\
&=&
\frac{1}{2(a_1a_3-a_2^2)}
\left(
  a_3 P_Q^2
- 2a_2 P_Q P_y
+ a_1 P_y^2
\right)
\nonumber\\
&+&
P_x Q
-\frac{1}{2}Q^2
+ V(x,y),
\ee
where the dependence on $P_x$ appears only in the linear term $P_x Q$.
Since there is no constraint which relates $P_x$
with the other canonical variables,
the Hamiltonian Eq. \eqref{ham_ng}
is not bounded from below,
which indicates the appearance of the Ostrogradsky instabilities.
We note that the above discussion cannot be applied
to the case of  $a_1a_3-a_2^2=0$,
which needs to considered separately.


\subsection{The degenerate case}

Second, we consider the {\it degenerate} case 
\be
\label{dege}
\frac{\partial^2 L_{p,2}}{\partial \dot{Q}^2}
\frac{\partial^2 L_{p,2}}{\partial \dot{y}^2}
-
\left(\frac{\partial^2 L_{p,2}}{\partial \dot{Q}\partial \dot{y}}\right)^2
=
a_1a_3-a_2^2
=0,
\ee
under which $P_Q$ and $P_q$
satisfy
\be
P_y-\frac{a_2}{a_1}P_Q
= 0.
\ee
Regarding
\be
\label{x1}
X_1:=
P_y-\frac{a_2}{a_1}P_Q\approx 0,
\ee
as the primary constraint,
the total Hamiltonian can be defined as  
\be
\label{toth}
{\tilde H}_p
&:=&
H_p
+\mu X_1
\nonumber\\
&=&
\frac{a_1P_y^2}{2a_2^2}
+P_x Q
-\frac{1}{2}Q^2
+V\left(x,y\right)
+
\mu X_1.
\ee
The time evolution of the primary constraint $X_1$
then
generates the secondary constraint
\be
\label{x2}
X_2
&:=&
\dot{X}_1
=\{X_1,{\tilde H}_p \}
\nonumber\\
&=&
-g_2x
-g_3 y
+
\frac{a_2}{a_1}  (P_x- Q)
\approx
0,
\ee
where we define the Poisson bracket
\be
\{U_1, U_2\} 
&:= &
\left(
\frac{\partial U_1}
        {\partial x}
\frac{\partial U_2}
       {\partial P_x}
-
\frac{\partial U_1}
        {\partial P_x}
\frac{\partial U_2}
       {\partial x}
\right)
\nonumber\\
&+&
\left(
\frac{\partial U_1}
       {\partial y}
\frac{\partial U_2}
       {\partial P_y}
-
\frac{\partial U_1}
        {\partial P_y}
\frac{\partial U_2}
        {\partial y}
\right)
\nonumber\\
&+&
\left(
\frac{\partial U_1}
        {\partial Q}
\frac{\partial U_2}
        {\partial P_Q}
-
\frac{\partial U_1}
        {\partial P_Q}
\frac{\partial U_2}
        {\partial Q}
\right).
\ee
The secondary constraint \eqref{x2}
relates $P_x$ to the other phase space variables
and 
all the terms linear in the momentum
are eliminated from the total Hamiltonian \eqref{toth}.
In other words, the Hamiltonian can be bounded from below.
Since
\be
\label{x1x2}
\{
X_1,X_2
\}
=
g_3
-
\frac{a_2^2}{a_1^2},
\ee
the time evolution of the secondary constraint $X_2$,
$\dot{X_2}=\{X_2, \tilde{H}_p\}\approx 0$,
fixes $\mu$
and generates no more constraint
for $g_3\neq \frac{a_2^2}{a_1^2}$.
From Eq. \eqref{x1x2},
we notie that 
the constraints  $X_1 \approx 0$ and $X_2\approx 0$,
Eqs. \eqref{x1} and \eqref{x2},
are of the second-class.
Thus, 
starting from the $6$-dimensional phase space $(x, y,Q,P_x,P_y, P_Q)$,
the $2$ second-class constraints
reduce the dimensionality of the phase space to be $4(=2\times2)$,
namely,
and hence
the correct  $2$ degrees of freedom are recovered
and the Ostrogradsky ghosts are eliminated.

\section{The limit to the quadratic-order DHOST theories}
\label{app_b}

In the limit of the scalar-tensor theories,
$A_\mu\to \nabla_\nu\phi$,
the quadratic-order extended vector-tensor theories
\eqref{gp} with Eq. \eqref{defc4}
reduce to the quadratic-order DHOST theories 
\cite{Langlois:2015cwa,Achour:2016rkg},
\begin{eqnarray}
\label{action2}
S
&=&
\int  d^4x\sqrt{-g}
\left[
f_0 (\X)
+
f_2\left({\X} \right){R}
\right.
\nonumber\\
&&
\left.
+{C}^{\mu\nu\rho\sigma}
\left({\nabla}_\mu \nabla_\nu \phi \right)
\left({\nabla}_\rho \nabla_\sigma \phi\right)
\right],
\end{eqnarray}
with
\begin{eqnarray}
\label{c_sca}
&&
C^{\mu\nu\rho\sigma}
\left(\nabla_\mu \nabla_\nu \phi\right) 
\left(\nabla_\rho \nabla_\sigma \phi\right)
\nonumber\\
&=&
 \alpha_1(\X)
  \left(\nabla_\mu \nabla_\nu \phi\right)
  \left(\nabla^\mu \nabla^\nu \phi\right)
+\alpha_2(\X)
   \left(\Box\phi\right)^2
\nonumber\\
&+&\alpha_3(\X)
\left(\nabla^\mu \phi\right)\left(\nabla^\nu \phi \right)
\left(\nabla_{\mu}\nabla_{\nu}\phi\right)
 \Box\phi
\nonumber\\
&+&
\alpha_4(\X)
\left(\nabla^\mu \phi\right)\left(\nabla^\nu \phi\right)
\left(\nabla_{\mu}\nabla_{\rho}\phi\right)
\left(\nabla_\nu \nabla^\rho \phi\right)
\nonumber\\
&+&\alpha_5(\X)
 \left[\left(\nabla^\mu \phi\right)\left(\nabla^\nu \phi\right) \nabla_{\mu}\nabla_{\nu}\phi\right]^2,
\end{eqnarray}
where we have defined 
\be
\label{defx}
\X:=g^{\mu\nu}
\nabla_\mu \phi
\nabla_\nu \phi,
\qquad 
\Box\phi
:=g^{\mu\nu}
\nabla_\mu \nabla_\nu \phi,
\ee
and $\nabla_\mu$ denotes the covariant derivative
associated with the metric $g_{\mu\nu}$.
The relevant degeneracy conditions
were obtained in Refs.
\cite{Langlois:2015cwa,Achour:2016rkg,BenAchour:2016fzp,Langlois:2018dxi}.

We classify our exact black hole solutions
into the two cases in terms of the constant value of $\X$;

\begin{itemize}

\item
Case-1; $\X_0=-q^2$.

``$\Lambda$'' is also attached
in the case that the solution is asymptotically de Sitter or anti-de Sitter. 

\item
Case-2; $\X_0\neq  -q^2$.

``$\Lambda$'' is also attached
in the case that the solution is asymptotically de Sitter or anti-de Sitter.

\end{itemize}
Accordingly,
in the limit of the scalar-tensor theories, 
the conditions Case 1-1, Case 2-1, Case 1-$\Lambda$, Case 2-1-$\Lambda$
reduce to 

\begin{itemize}

\item{Case 1}
\begin{subequations}
\be
\label{case11x}
f_0&=&f_{0,\X}=0,
\\
  \alpha_2
&=&-\alpha_1,
\\
\alpha_{2,\X}
&=&
-\alpha_{1,\X}.
\ee
\end{subequations}

\item{Case 2}
\begin{subequations}
\label{case21x}
\be
f_0&=&f_{0,\X}=0,
\\ 
\alpha_1
&=&
\alpha_2=0,
\\
\alpha_{2,\X}
&=&-\alpha_{1,\X},
\\
\alpha_3
&=&
-2\alpha_{1,\X}.
\ee
\end{subequations}

\item{Case 1-$\Lambda$}
\begin{subequations}
\label{case13x}
\be
f_0
&=&
-2\Lambda
\left(
  f_2
+q^2\alpha_1
\right),
\\
f_{0,\X}
&=&
\Lambda
\left[
\alpha_1
+
\frac{3}{2}q^2
\alpha_3
-
4f_{2,\X}
+
2q^2\alpha_{1,\X}
\right],
\\
\alpha_2
&=&-\alpha_1,
\\
\alpha_{2,\X}
&=&
-\alpha_{1,\X}.
\ee
\end{subequations}

\item{Case 2-$\Lambda$}
\begin{subequations}
\label{case23x}
\be
f_0
&=&
-2\Lambda f_2,
\nonumber\\
f_{0,\X}
&=&
\Lambda
\left(
-4f_{2,\X}
+\X_0 \alpha_{1,\X}
\right),
\nonumber
\\
\alpha_1
&=&
\alpha_2
=0,
\nonumber\\
\alpha_{2,\X}
&=&
-\alpha_{1,\X},
\nonumber\\
\alpha_3
&=&
-2\alpha_{1,\X}.
\ee
\end{subequations}

\end{itemize}
These conditions exactly agree with
those in  
``Case 1'', ``Case 2'', ``Case 1-$\Lambda$'', and ``Case 2-$\Lambda$''
obtained in Ref. \cite{dhostbh2} (see also Refs. \cite{dhostbh1,Takahashi:2020hso}), respectively.

\section{The Reissner-Nordstr\"{o}m solution in the Einstein-Maxwell theory}
\label{app_c}

In this Appendix,
we illustrate the derivation of the black hole solutions
in the Einstein-Maxwell theory
\be
\label{em}
S=
\int d^4 x\sqrt{-g}
\left[
\frac{M_p^2}{2}R
-\frac{1}{4}F^{\mu\nu}F_{\mu\nu}
\right].
\ee
We assume the ansatz for the metric and vector field \eqref{ansatz}
where without loss of generality we may set $A_r=0$.
Substituting Eq. \eqref{ansatz} 
into the Einstein-Maxwell action \eqref{em},
\begin{widetext}
\be
S=\int dt dr
\frac{\pi} {f^2}
\sqrt{\frac{f}{h}}
\left\{
  M_p^2r^2 h f'^2
-4M_p^2 h^2
  \left(-1+h+rh'\right)
+rf 
\left[
-M_p^2 rf'h'
+2h
\left(
rA_t'^2 
-M_p^2 \left(2f'+rf''\right)
\right)
\right]
\right\}.
\ee
\end{widetext}
Varying the action with respect to $f$, $h$, and $A_t$,
we obtain 
\begin{subequations}
\begin{eqnarray}
\label{aa1}
&&
-r^2 h A_t'^2
+2M_p^2 f\left(-1+h+rh'\right)
=0,
\\
\label{aa2}
&&
2M_p^2 f \left(-1+h\right)
+rh \left(r A_t'^2+2M_p^2 f'\right)
=0,
\\
\label{aa3}
&&
rh A_t' f'
-f
\left[
rA_t' h'
+2h \left(rA_t''+2A_t'\right)
\right]
=0.
\end{eqnarray}
\end{subequations}
By integrating Eq. \eqref{aa1},
\be
\label{qq1}
A_t'(r)=-\frac{Q}{r^2} \sqrt{\frac{f(r)}{h(r)}},
\ee
where $Q$ is an integration constant.
Substutiting it into Eq,. \eqref{aa1}, we obtain 
\be
\frac{Q^2}{r^2}
+2M_p^2 
\left(
rh'+h-1
\right)
=0,
\ee
which can be integrated as 
\be
\label{qq2}
h(r)
=1-\frac{2M}{r}+\frac{Q^2}{2M_p^2 r^2},
\ee
where $M$ is an integration constant.
Finally, substituting Eqs. \eqref{qq1} and \eqref{qq2} 
into Eq. \eqref{aa2},
\be
&&
2\left(Q^2-2MM_p^2 r\right)f(r)
\nonumber\\
&+&
r
\left(
Q^2+2M_p^2 r(r-2M)
\right)
f'(r)
=0,
\ee
which can be integrated as 
\be
f(r)=c_1 h(r),
\ee
where $c$ denotes an integrating constant
corresponding to the rescaling of the time coordinate. 
Without loss of generality, we can set $c_1=1$
and then reproduce the Reissner-Nordstr\"{o}m solution
\begin{subequations}
\label{rn_em}
\begin{eqnarray}
f(r)&=& h(r)=1-\frac{2M}{r}+\frac{Q^2}{2M_p^2 r^2},
\\
A_t'(r)&=&-\frac{Q}{r^2},
\end{eqnarray}
\end{subequations}
where the integration constants $M$ and $Q$
physically
represent the mass and electric charge of the black hole.

\bibliography{disformal_refs}
\end{document}